\documentclass{iau} 
\usepackage{graphicx}
\usepackage{amsmath}
\usepackage{bm}

\title[Polarised emission from black holes] 
{Modelling the polarised emission from black holes on event horizon-scales}

\author[Younsi, Porth, Mizuno, Fromm \& Olivares]   
{Ziri Younsi$^{1,2}$, 
Oliver Porth$^{1}$, 
Yosuke Mizuno$^{1}$, 
Christian M.~Fromm$^{1,3}$
\& Hector Olivares$^{1}$}

\affiliation{$^1$Sterrenkundig Instituut, University of Utrecht, \\ Postbus 80000,
NL-3508TA, Utrecht, the Netherlands \\ email: {\tt m.lugaro@phys.uu.nl} \\[\affilskip]
$^2$Dept. of Astronomy \& Space Physics, Uppsala University, \\ Box
515, SE-75120 Uppsala, Sweden \\email: {\tt hoefner@astro.uu.se}}

\affiliation{$^1$Institut f\"ur Theoretische Physik, Max-von-Laue-Stra{\ss}e 1, 
D-60438 Frankfurt am Main, Germany \\ email: {\tt younsi@itp.uni-frankfurt.de} \\[\affilskip]
$^{2}$Mullard Space Science Laboratory, University College London,
Holmbury St.\,Mary, Dorking, Surrey RH5 6NT, UK \\ email: {\tt z.younsi@ucl.ac.uk} \\[\affilskip]
$^{3}$Max-Planck-Institut f\"ur Radioastronomie, Auf dem H\"ugel 69,
D-53121 Bonn, Germany}

\pubyear{2018}
\volume{342}  
\jname{Perseus in Sicily: From Black Hole to Cluster Outskirts}
\editors{K.~Asada, E.~de G.~dal Pino, H.~Nagai, R.~Nemmen \& M.~Giroletti}
\begin{document}

\maketitle

\begin{abstract}
Upcoming VLBI observations will resolve nearby supermassive
black holes, most notably Sagittarius A* and M87, on event horizon-scales.
Recent observations of Sagittarius A* with the Event Horizon Telescope have
revealed horizon-scale structure.
Accordingly, the detection and measurement of the back hole ``shadow"
is expected to enable the existence of astrophysical black holes to be
verified directly.
Although the theoretical description of the shadow is straightforward,
its observational appearance is largely determined by the
properties of the surrounding accretion flow, which is highly turbulent.
We introduce a new polarised general-relativistic radiative
transfer code, \texttt{BHOSS}, which accurately solves the equations of
polarised radiative transfer in arbitrary strong-gravity environments,
providing physically-realistic images of astrophysical black holes on
event horizon-scales, as well as also providing insight into the
fundamental properties and nature of the surrounding accretion flow
environment.
\end{abstract}

\firstsection 
\section{Introduction}
It is widely believed that all galaxies host a supermassive black hole (SMBH)
at their center.
Advances in very-long-baseline-interferometry (VLBI) have enabled the
Event Horizon Telescope Collaboration (EHTC) to image (with event horizon-scale
resolution) the nearby SMBHs Sagittarius A* (Sgr A*) and M87,
with the first observational results expected soon (e.g., \cite{Doeleman08}, \cite{Goddi17}).
When observing an astrophysical black hole, theorists anticipate seeing a
``shadow", the silhouette of the unstable photon region surrounding the event horizon
(e.g., \cite{Cunningham73}, \cite{Grenzebach14}, \cite{Younsi16}).
Gravity near the event horizon is so strong that spacetime is
curved and light rays no longer travel along straight lines.
Moreover, the surrounding environment is hot, magnetised and
turbulent, serving to obscure and modify the shadow's observed properties.
As such, both the equations of motion for light rays and the radiative transfer
equation must be solved in full general relativity.

We present results from our new general-relativistic radiative transfer
(GRRT) code, \texttt{BHOSS} (Younsi \etal\ 2018, in prep.), which solves the equations
of polarised radiative transfer for arbitrary spacetime metrics
(e.g., \cite{Younsi16}, \cite{Mizuno18}), and is fully coupled to many three-dimensional
general-relativistic magnetohydrodynamical (GRMHD) codes,
including \texttt{BHAC} (\cite{Porth17}) and \texttt{HARM} (\cite{Gammie03}).
\section{General-relativistic ray-tracing and radiative transfer}
\begin{figure}[t]
\begin{center}
\includegraphics[trim={1.1mm 0 0.9mm 0},clip,width=0.32\linewidth]{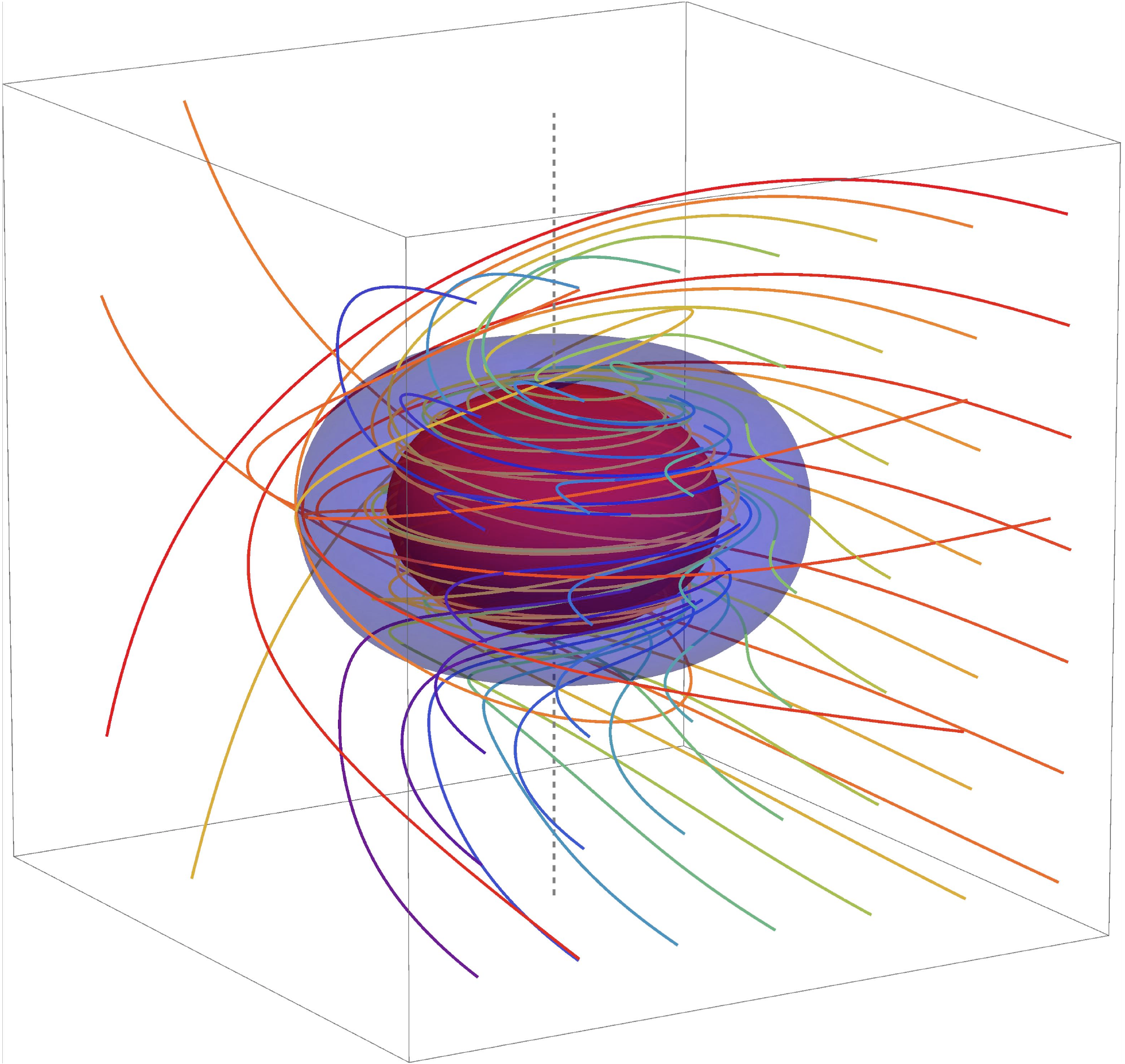} 
\includegraphics[width=0.32\linewidth]{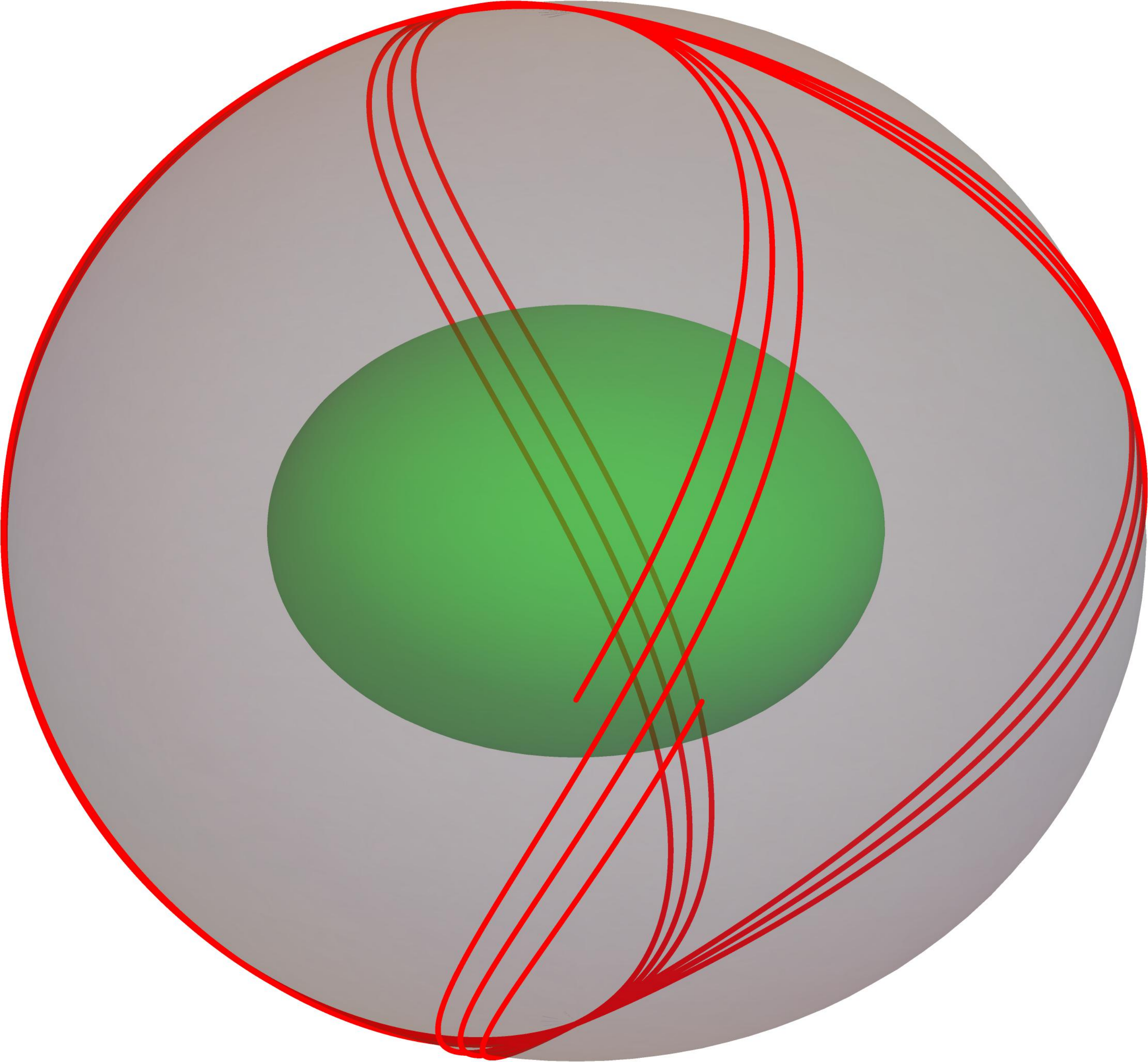} 
\includegraphics[width=0.32\linewidth]{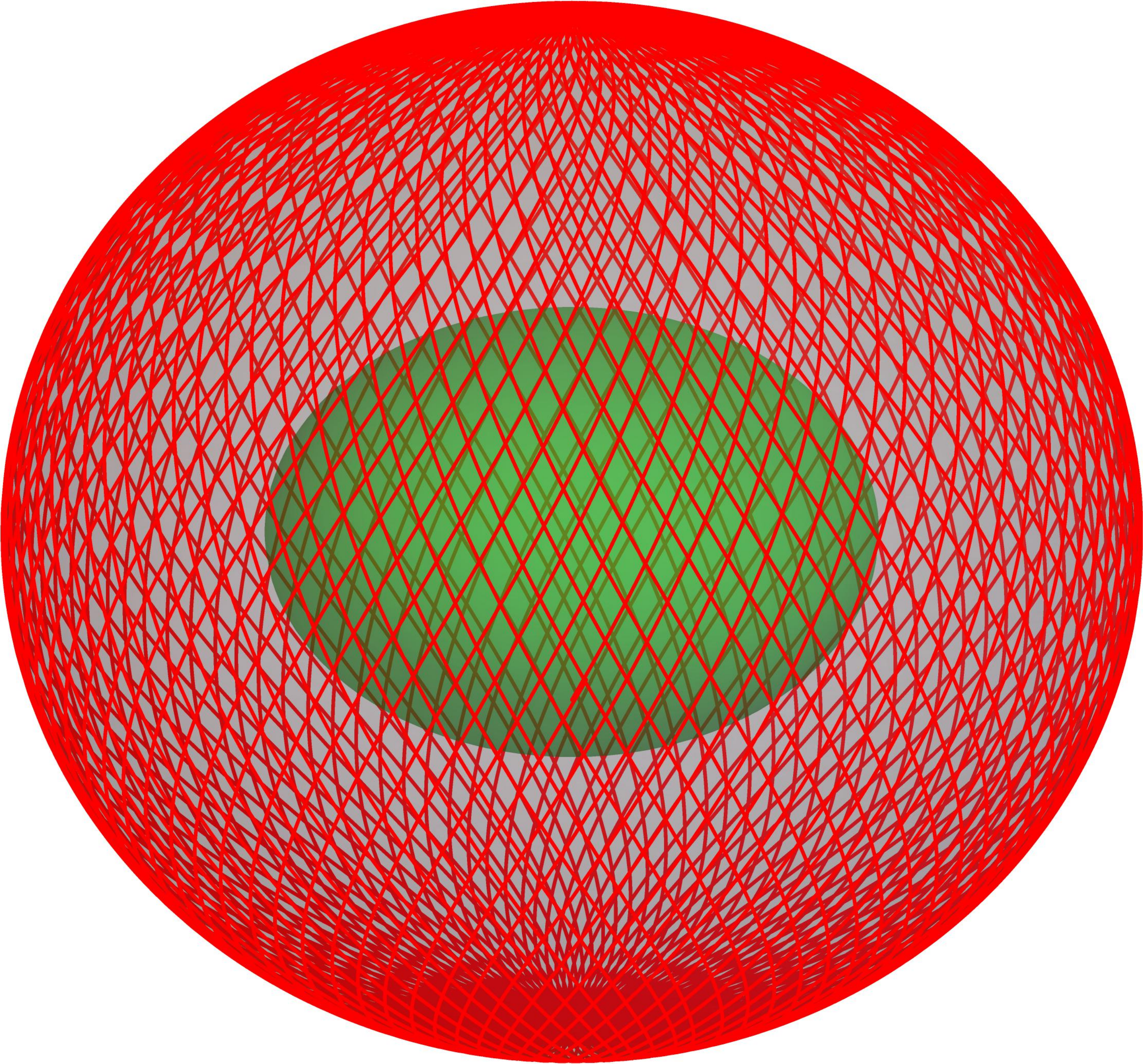} 
\caption{{\it Photon geodesics around an extremal Kerr black hole}.
\textbf{Left panel:} $7\times 7$ pixel observer grid of rays with event horizon in red and ergo-region in blue.
\textbf{Middle panel:} $6\pi$ longitudinal oscillations of a spherical photon orbit (event horizon in green).
\textbf{Right panel:} $100\pi$ longitudinal oscillations of the spherical photon orbit.
The orbit is stable for long integration times.}
\label{fig1}
\end{center}
\end{figure}

{\it Ray-tracing}.
In order to construct an image of a black hole, the paths of light rays (photons)
comprising an image are ray-traced by solving the geodesic equations of motion:
\begin{eqnarray}
\frac{{\rm d} x^{\alpha}}{{\rm d} \lambda} &=& k^{\alpha} \,, \\
\frac{{\rm d} k^{\alpha}}{{\rm d} \lambda} &=& -\Gamma^{\alpha}_{\phantom{\alpha}\mu\nu}
k^{\mu}k^{\nu} \,,
\end{eqnarray}
where $x^{\alpha}$ and $k^{\alpha}$ are, respectively, the photon position and 4-momentum,
$\Gamma^{\alpha}_{\phantom{\alpha}\mu\nu}$ denote the Christoffel symbols, $\lambda$ is the
affine parameter and Greek indices range from $0$--$3$ throughout.
Appropriate initial conditions may be determined using a pre-defined observer ``camera"
(see, e.g., \cite{Younsi16} and left panel of Fig.~\ref{fig1}).
Tensor index raising and lowering operations are performed with the metric tensor, $g_{\mu\nu}$,
e.g., $k_{\mu}=g_{\mu\nu}k^{\nu}$.
Geodesics in any spacetime may be calculated by specifying the appropriate $g_{\mu\nu}$.

Numerical errors in geodesic integration grow over time, and are minimised in \texttt{BHOSS}
by using several different high-order integration schemes.
A robust test of stability is the long-term integration of spherical photon orbits
(e.g., \cite{Teo03}, \cite{Chan17}), which are notoriously challenging to maintain for
long times but readily maintained in \texttt{BHOSS}
(see middle and right panels of Fig.~\ref{fig1}).

{\it Polarised radiative transfer}.
Along each ray (image pixel) the intensity and polarisation properties are calculated
from the equations of polarised radiative transfer.
The observer's polarisation basis can be represented by parallel and perpendicular
4-vectors of this basis, $f^{\alpha}_{\left(\parallel\right)}$ and $f^{\alpha}_{\left(\perp\right)}$,
which are orthogonal both to each other and to $k^{\alpha}$, and solved as:
\begin{equation}
\frac{{\rm d} f^{\alpha}}{{\rm d} \lambda} = -\Gamma^{\alpha}_{\phantom{\alpha}\mu\nu}
k^{\mu} f^{\nu} \,.
\end{equation}
Defining the Lorentz-invariant Stokes vector as
$\bm{\mathcal{S}}:=\left( \mathcal{I}, \mathcal{Q}, \mathcal{U}, \mathcal{V}\right)^{\rm T}$,
where $\bm{\mathcal{S}}\equiv {\bf S}/\nu^{3}$ and
${\bf S}:=\left( I, Q, U, V \right)^{\rm T}$, the full polarised GRRT equation may be written as:
\begin{equation}
\frac{{\rm d}}{{\rm d} \lambda}
\begin{pmatrix}
\mathcal{I} \\ 
\mathcal{Q} \\ 
\mathcal{U} \\ 
\mathcal{V}
\end{pmatrix}
=-k_{\mu}u^{\mu}\left\{R\left(\chi \right)
\begin{pmatrix}
\overline{\varepsilon}_{I} \\ 
\overline{\varepsilon}_{Q} \\ 
0 \\ 
\overline{\varepsilon}_{V}
\end{pmatrix}
-\left[R\left(\chi \right)
\begin{pmatrix}
\overline{\alpha}_{I} & \overline{\alpha}_{Q} & 0 & \overline{\alpha}_{V} \\ 
\overline{\alpha}_{Q} & \overline{\alpha}_{I} & \overline{\rho}_{V} & 0 \\ 
0 & -\overline{\rho}_{V} & \overline{\alpha}_{I} & \overline{\rho}_{Q} \\ 
\overline{\alpha}_{V} & 0 & -\overline{\rho}_{Q} & \overline{\alpha}_{I}
\end{pmatrix}
R\left(-\chi \right) \right]
\begin{pmatrix}
\mathcal{I} \\ 
\mathcal{Q} \\ 
\mathcal{U} \\ 
\mathcal{V}
\end{pmatrix}
\right\} \,,
\label{PGRRT}
\end{equation}
where $\overline{\varepsilon}_{i}$ are the invariant emissivities,
$\overline{\alpha}_{i}$ the invariant absoprtivities and $\overline{\rho}_{i}$ are
the invariant Faraday rotation and conversion coefficients.
Invariant (barred) quantities are related to their standard (fluid rest frame) counterparts as
$\overline{\varepsilon}_{i}=\varepsilon_{i}/\nu^{2}$,
$\overline{\alpha}_{i} = \nu \alpha_{i}$ and $\overline{\rho}_{i} = \nu \rho_{i}$,
where $\nu$ is the frequency (see \cite{Younsi12}).
The 4-velocity of the fluid, e.g., an accretion flow around a black hole, is denoted by $u^{\mu}$
and $R\left( \chi \right)$ is a rotation matrix where $\chi\in [-\pi,\pi]$ is the angle between the polarisation
basis and the plasma magnetic field, also termed the electric vector position angle (EVPA).
Note that $\varepsilon_{U}=\alpha_{U}=\rho_{U}=0$ is fixed by rotating to a frame aligning
the magnetic field with $U$.
Equation \eqref{PGRRT} is integrated along each photon geodesic, returning the Stokes
parameters, ${\bf S}$.
\section{Preliminary tests and future work}
\begin{figure}[t]
\begin{center}\hspace{2mm}
\includegraphics[trim={0 9mm 0 0},clip,width=0.46\linewidth]{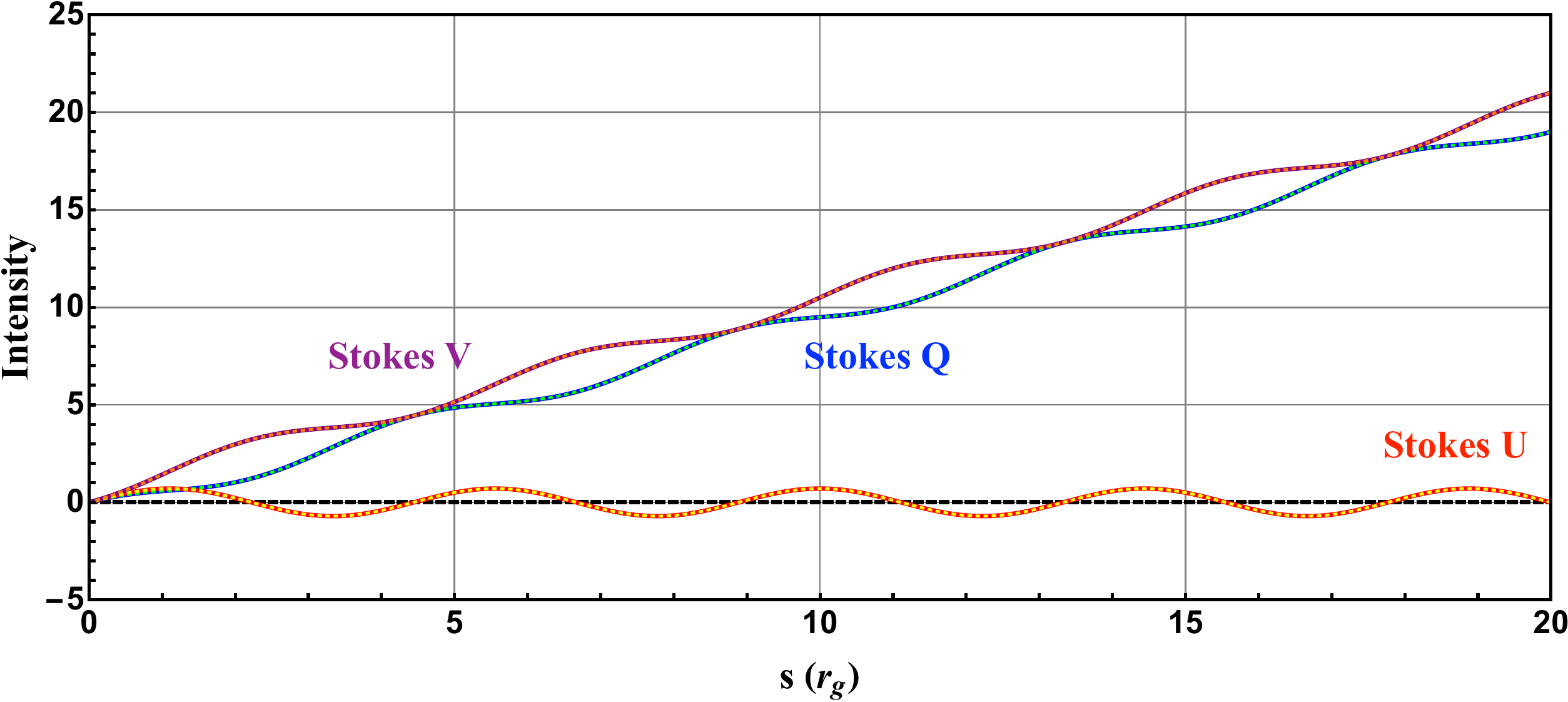}\hspace{2mm}
\includegraphics[trim={7mm 9mm 0 0},clip,width=0.447\linewidth]{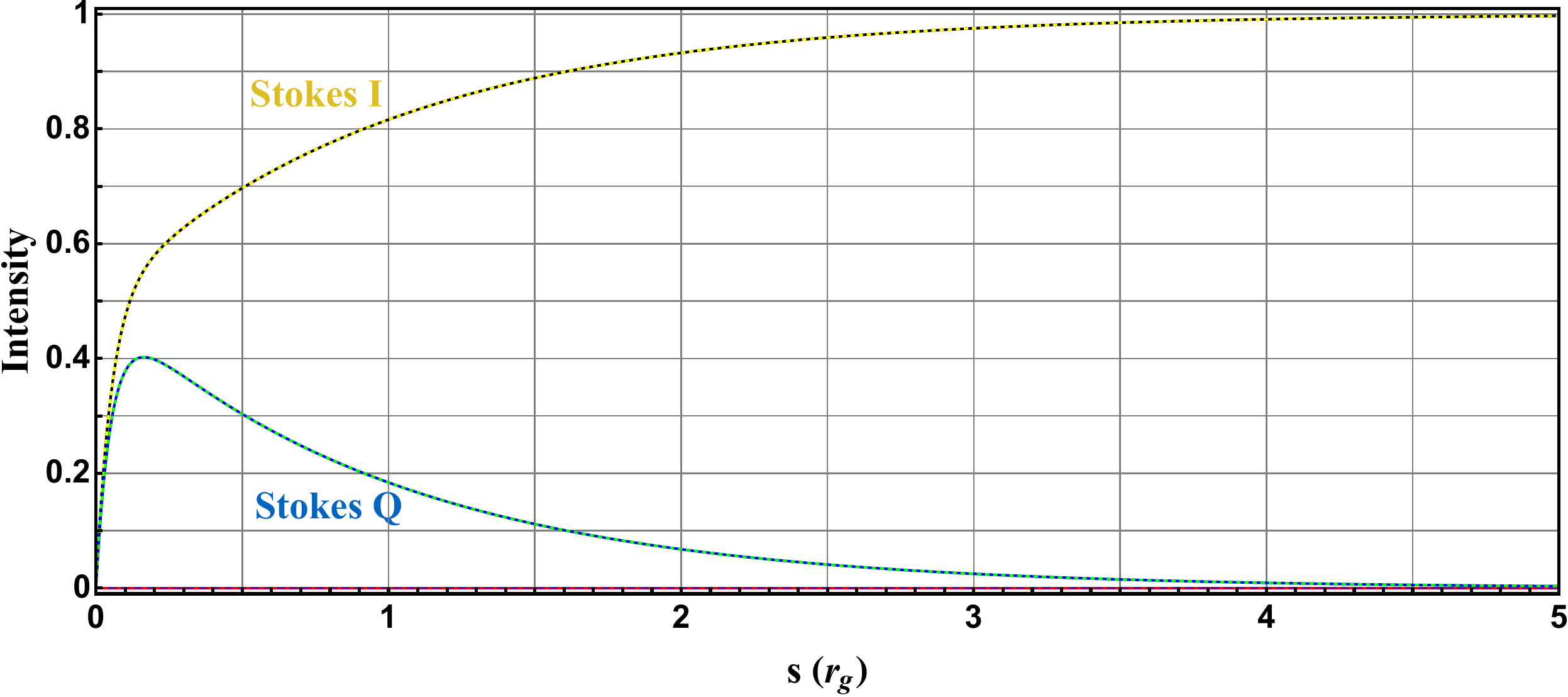} \\
\includegraphics[width=0.49\linewidth]{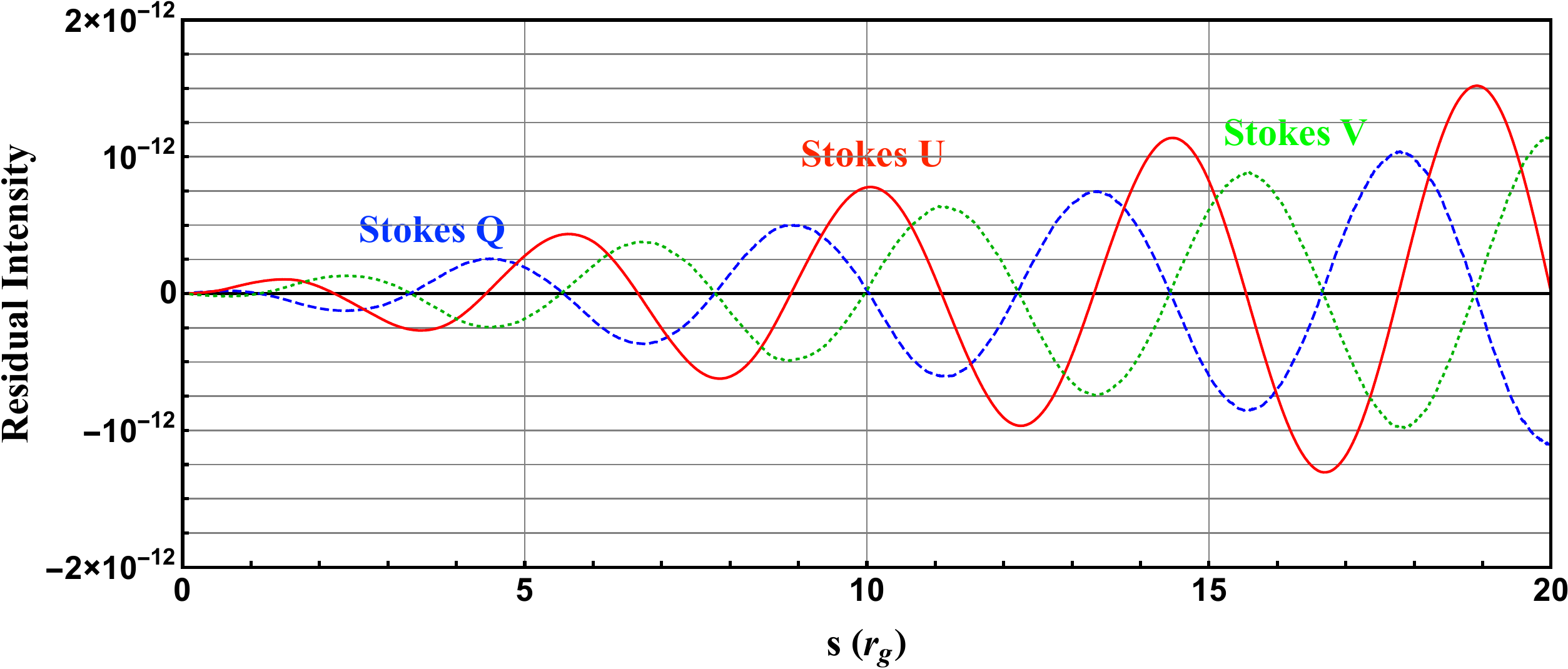} 
\includegraphics[trim={7mm 0 0 0},clip,width=0.47\linewidth]{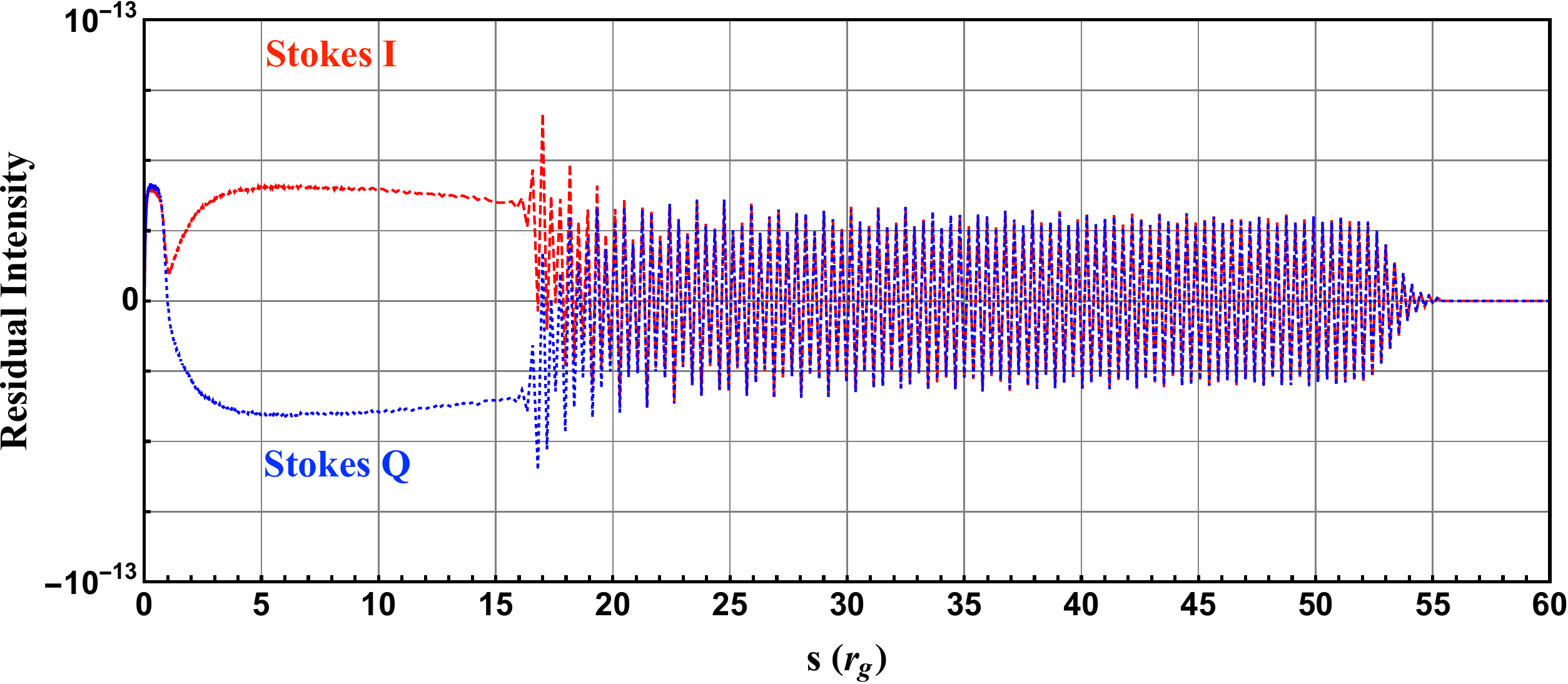} 
\caption{{\it Polarised GRRT tests I}.
\textbf{Left column:} pure polarised emission in Stokes $(Q,U,V)$, $I=0$ ({\it top panel}) and residuals
compared to analytic solution ({\it bottom panel}).
\textbf{Right column:} pure emission and absorption in Stokes $(I,Q)$, $U=V=0$ ({\it top panel})
and residuals compared to analytic solution ({\it bottom panel}).
Analytic solutions are plotted as dashed lines in upper panels.}
\label{fig2}
\end{center}
\end{figure}
\begin{figure}[t]
\begin{center}\hspace{2mm}
\includegraphics[trim={0 0mm 0 0},clip,width=0.47\linewidth]{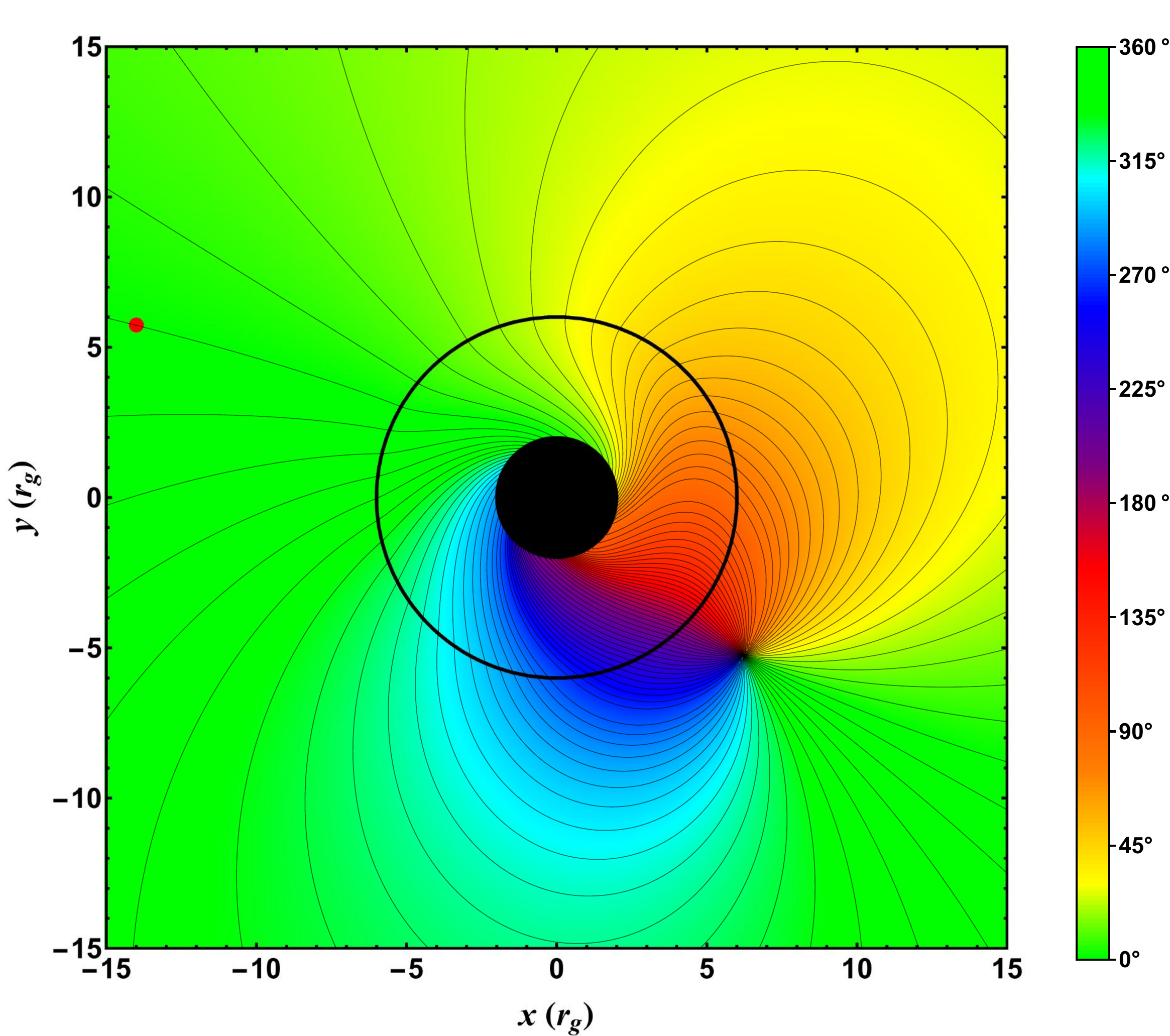}
\includegraphics[trim={0mm 0mm 0 0},clip,width=0.485\linewidth]{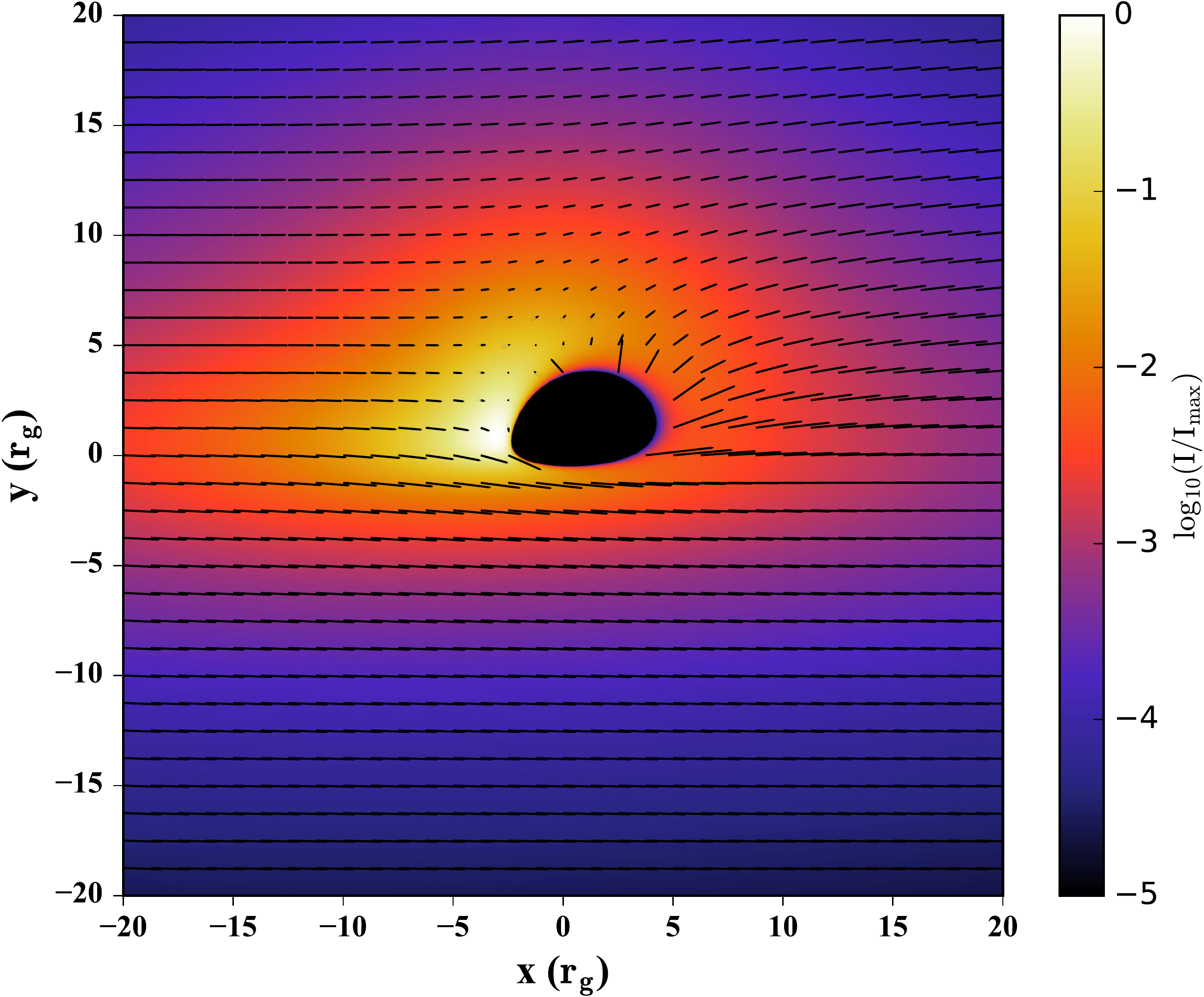}
\caption{{\it Polarised GRRT tests II}.
\textbf{Left panel:} EVPA contour map ($\widetilde{\chi}\equiv\chi + \pi$) of a Keplerian accretion
disk around a Schwarzschild BH as viewed at $30^{\circ}$.
The black ring denotes the innermost stable circular orbit (ISCO), wherein the disk material is in free-fall.
The red dot corresponds to the $\widetilde{\chi}=0^{\circ}$ contour.
\textbf{Right panel:} Novikov-Thorne accretion disk with electron scattering atmosphere around
an extremal Kerr black hole.
Colour denotes the total intensity integrated over the range $0.1$--$10$~keV,
$\dot{M}=0.1~M_{\rm Edd}$, $M=10~M_{\odot}$ and viewed at $75^{\circ}$.
Orientation of black ticks indicates the polarisation direction, and the length of each tick
denotes the polarisation degree ($\sim5$\% at bottom of panel).
Significant depolarisation and enhancement can be seen.}
\label{fig3}
\end{center}
\end{figure}
We present preliminary tests of the above polarised GRRT framework in \texttt{BHOSS}.
Several polarised GRRT codes exist in the literature (\cite{Broderick03ab}, \cite{Schnittman10},
\cite{Shcherbakov11}, \cite{Dexter16}, \cite{Moscibrodzka18}, \cite{Pihajoki18}) which
contain several standard tests.
Figure \ref{fig2} shows two ray transport tests (see \cite{Dexter16}): (i) transport of pure
polarised emission with Faraday rotation and conversion, where
$\epsilon_{Q}=\epsilon_{U}=\epsilon_{V}=\rho_{Q}=\rho_{V}=1$, $\rho_{U}=0$,
and (ii) pure emission and absorption in Stokes I and Q, with
$\varepsilon_{I}=\alpha_{I}=10$, $\varepsilon_{Q}=\alpha_{Q}=9$.
The numerical integration performs well and, as expected, Stokes $I$ is zero to machine
precision in test (i) and Stokes $U$ and $V$ are similarly zero in test (ii).

In Fig.~\ref{fig3} we present more detailed polarised GRRT imaging tests.
The left panel shows the EVPA map of a thin Keplerian accretion disk around
a Schwarzschild black hole, and is in good agreement with that of Fig.~$3$ in \cite{Dovciak08}.
The right panel presents the image of a Novikov-Thorne accretion disk with an
electron scattering atmosphere (see \cite{Schnittman10} for further details).
The obtained image is in good qualitative agreement with previous results obtained by
\cite{Schnittman10} and \cite{Dexter16}.

We have presented a numerical GRRT code capable of integrating the equations
of polarised GRRT to high accuracy in arbitrary spacetime geometries, demonstrating
excellent agreement with previous results and analytic solutions in the literature.
Since \texttt{BHOSS} already performs GRRT in time-dependent optically-thin and optically-thick
GRMHD backgrounds, and for arbitrary spacetime geometries, 
we will next perform polarised GRRT calculations on GRMHD simulation data, determining time-variable
polarised emission from accreting black holes and providing meaningful theoretical predictions
to be compared with upcoming and future observing campaigns.

\begin{acknowledgements}
We thank Luciano Rezzolla, Hung-Yi Pu and Thomas Bronzwaer for helpful input.
Support comes from the ERC Synergy Grant ``BlackHoleCam - Imaging the
Event Horizon of Black Holes'' (Grant 610058).
ZY is supported by a Leverhulme Trust Early Career Fellowship.
This research has made use of NASA's Astrophysics Data System.
\end{acknowledgements}

\end{document}